\documentstyle[aps,prd,graphicx]{revtex}

\begin{document}

\preprint{ULB-TH-00/22}

\title{Neutrino suppression and extra dimensions: A minimal model}

\author{N.~Cosme, \footnote{E-mail address: ncosme@ulb.ac.be}
J.-M.~Fr\`ere, 
Y.~Gouverneur,\footnote{E-mail address:ygouvern@ulb.ac.be}
F.-S.~Ling, \footnote{E-mail address: fsling@ulb.ac.be}
D.~Monderen, \footnote{E-mail address: dmondere@ulb.ac.be}
and  V.~Van~Elewyck \footnote{E-mail address: vvelewyc@ulb.ac.be}
}

\address{Service de Physique Th\'eorique, CP225, 
Universit\'e Libre de Bruxelles, Bvd du Triomphe, B-1050 Brussels, Belgium}

\maketitle

\begin{abstract}
We study flavor neutrinos confined to our four-dimensional world coupled to one
"bulk" state, i.e., a Kaluza-Klein tower. We discuss the spatial development of the 
neutrino disappearance, the possibility of resurgence and the effective flavor 
transitions induced in this mechanism. We show that even a simple model can produce
an energy independent suppression at large distances, and relate this to 
experimental data.

\end{abstract}

\section{Introduction}
\label{s:intro}

The possibility of "large extra dimensions", i.e., with (at least) one
compactification radius close to the current validity limit of Newton's law of gravitation
($\sim$ 1 mm), raises considerable interest.
It could also solve the hierarchy problem by providing us with a new fundamental scale
at an energy possibly as small as 1 TeV \cite{led,antoniadis}. 

Neutrino physics is a favorite area to
study this possibility. A right-handed, sterile neutrino does not experience any of the gauge interactions 
that require the confinement of the other standard model particles to our four-dimensional brane;
it is thus, other than the graviton, an ideal tool to probe the "bulk" of space. 
Unconventional patterns of neutrino masses and oscillations arise, taking advantage of 
the possibility that the flavor neutrinos confined to our space can now interact 
with the bulk states that appear to us (due to compactification of the extra
dimensions) as so-called Kaluza-Klein towers of states \cite{lednu,gherghetta}. 
Recent works \cite{dvali,ramond,lorenzana,creminelli,mohapatra} have shown that,
at least partially, it is possible to accommodate experimental constraints 
on neutrinos within this setup.

 We explore further possibilities, focusing on the unique properties of the
model. Quite specifically, neutrinos in this scheme can "escape" for part of the time to extra dimensions,
resulting in a reduced average probability of detection in our world. While similar in a way to a fast unresolved
oscillation between flavor and sterile neutrinos, this differs both by the time development of the effect and
by the possible depth of the suppression.

In Sec.~\ref{s:sec2}, we address the analytical aspects of neutrino oscillations in the
simplest toy model of one flavor neutrino coupling to one massless bulk fermion, i.e., to its tower
of states. We recall the main equations and reinvestigate the neutrino survival probability in
order to obtain its correct behavior at large $L/E$. We review the experimental constraints for
both $\nu _e$ and  $\nu _{\mu}$ and inquire whether this toy model can accommodate them for
$\nu _e$ or for $\nu _{\mu}$. We show that the MSW matter effects in the Sun are inescapable for
the  $\nu _e$. However, as the large $L/E$ behavior of the survival probability leads naturally to
an energy independent spectrum, we further explore the possibility of accounting for the solar
neutrino data by a global suppression free from MSW effects. This can be achieved by extending
the toy model to include a second active neutrino.

In Sec.~\ref{s:sec3}, we thus propose an enlarged model that includes two
generations of flavor neutrinos, both coupling to the same bulk fermion. 
In this scheme, partial disappearance of both flavor neutrinos as well as oscillations 
between them become possible. We find the region of parameters that solve the
electronic neutrino problem, and then show that the $\nu _\mu$ constraints, except the LSND
result, can simultaneously be accounted for. 

\section{One neutrino coupled to one bulk fermion.}
\label{s:sec2}

The simplest model is constituted by one left handed neutrino $\nu _1$ (either of $
\nu _e$,$\nu _\mu $,$\nu _\tau $) which lives in our 3+1 dimensional
world coupled with one singlet bulk massless fermion field. Since the
latter lives in all dimensions, from our world's point of view, it
appears after compactification as a Kaluza-Klein tower, i.e., an infinite
number of four-dimensional spinors. Already in five dimensions, Dirac spinors
necessarily involve left and right-handed states (chirality is defined here in 3+1
dimensions); both will thus be present in the Kaluza-Klein tower.

\subsection{Basic relations}

The analysis presented here is based on a reduction of the theory from 4+1
to 3+1 dimensions. However, to guarantee a low scale for the unification of
gravity with all forces, more extra dimensions are needed. We will assume
that their compactification radii are small enough that they do not affect
the analysis. The pattern is now well established (see, e.g.,~\cite{dvali,creminelli})
and we will only recall the basic equations and results.

The action used is the following : 
\begin{equation}
S=\int d^4x\,dy\;\overline{\Psi} i\Gamma _A\partial ^A\Psi +\int d^4x\{\overline{\nu} _1 
i\gamma _\mu \partial ^\mu \nu _1+\lambda \overline{\nu} _1\Psi (x^\mu ,y=0)H(x^\mu
)+{\rm H.c.}\} ,
\end{equation}

\noindent where $A=0,...,4$ and $x^4=y$ is the extra dimension. 
The Yukawa coupling between the usual Higgs scalar, 
the weak eigenstate neutrino $\nu _1$, and the bulk
fermion operates at $y=0$, which is the 3+1
dimensional brane of our world.

The fifth dimension is taken to be a circle of radius $R$. As usual, the
bulk fermion $\Psi $ is expanded in eigenmodes. 
One then integrates over the fifth dimension.
Eventually, one has to diagonalize the mass matrix (eigenvalues noted 
$\lambda _n$) and write the neutrino in terms of the mass eigenstates 
\begin{eqnarray}
\label{valprop}
\lambda _n &=&\pi \xi ^2\cot (\pi \;\lambda _n) ,\\
\left| \nu _1\right\rangle &=&\sum _{n=0}^{\infty }U_{0n}\left| \nu
_{\lambda _n}\right\rangle \nonumber ,\\
\label{uon}
\left( U_{0n}\right) ^2 &=&\frac 2{1+\pi^2 \xi ^2+ \frac{\lambda _n ^2}{\xi ^2}},
\end{eqnarray}
\noindent where $\xi \equiv \frac m{1/R}$ 
measures the strength of the Yukawa coupling\footnote{Another convention introduces a 
$\sqrt{2}$ factor, as in~\cite{dvali,lorenzana}.} .

The survival amplitude $A_{\nu _1\nu _1}$and the survival probability 
$P_{\nu _1\nu _1}$ are given by 
\begin{eqnarray}
\label{anu1nu1}
A_{\nu _1\nu _1} &=&\sum _{n=0}^{\infty }\left( U_{0n}\right)
^2e^{\,i\left( \,\lambda _n\right) ^2x} ,\\
\label{pnu1nu1}
P_{\nu _1\nu _1} &=&\sum _{n=0}^{\infty }\left( U_{0n}\right) ^4+\sum
\sum _{n \neq m}\left( U_{0n}\right) ^2\left( U_{0m}\right) ^2\cos
\left[ \left( \left( \,\lambda _n\right) ^2-\left( \,\lambda _m\right)
^2\right) x\right],
\end{eqnarray}
where 
\begin{eqnarray}
x = \frac{L}{2ER^2} \approx 10^{-7}\frac{\left( L/{\rm km} \right) }{(E/{\rm GeV})(R/{\rm mm})^2}.
\end{eqnarray}
\begin{center}
\begin{figure}[h]
\centerline{\includegraphics{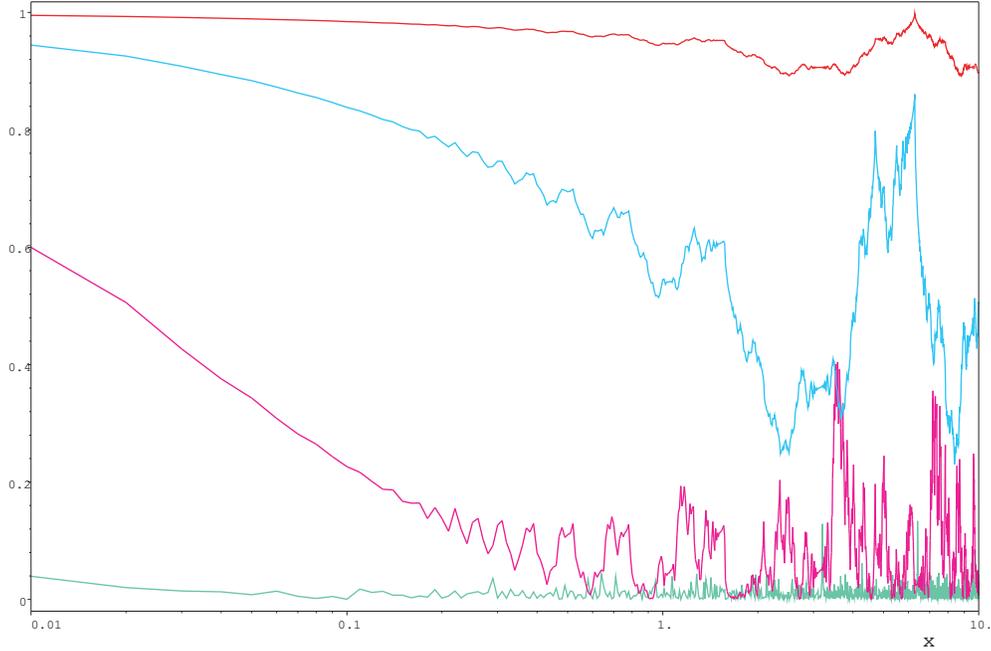}}
\caption{Survival probability for different values of $\xi$ 
($\xi=1/10, 1/3, 1, 3$ from top to bottom), as functions of $x=\frac{L}{2ER^2}$.}
\label{fig:prob}
\vspace{2pt}
\end{figure}
\end{center}
The main goal of the next section is to study the mathematical properties of
the survival probability and then to compare with experimental data.

\subsection{Behavior of survival amplitude and probability}

We give here a succinct description of the survival amplitude and probability.
Our main goal is to understand their behavior at large $x$, i.e.,
at large $L / E$ ratio.

We first discuss the eigenvalues of Eq.~(\ref{valprop}). They can be approached as 
follows by developing the cotangent:
\begin{eqnarray}
\label{valpropapprox}
\lambda _0 \approx & \frac{\xi}{\sqrt{1+\frac{\pi ^2 \xi ^2}{3}}}, & \\
\lambda _n \approx & n + \frac{\xi^2}{n} + \left( \frac{\pi^2 \xi^2}{3} +1\right) 
\frac{\xi^4}{n^3} ,&
{\rm for \,} n \gg \xi^2 ,\\
\lambda _n \approx &  n+\frac{1}{2}  + \frac{n+\frac{1}{2}}{1+\pi^2 \xi^2} - \frac{\pi^2 \xi^4
}{\left( 1 + \pi^2 \xi^2 \right) ^4} \left(
n+ \frac{1}{2} \right) ^3 ,& {\rm for \,} n \ll \xi^2.
\end{eqnarray}

\begin{center}
\begin{figure}[t]
\centerline{\includegraphics{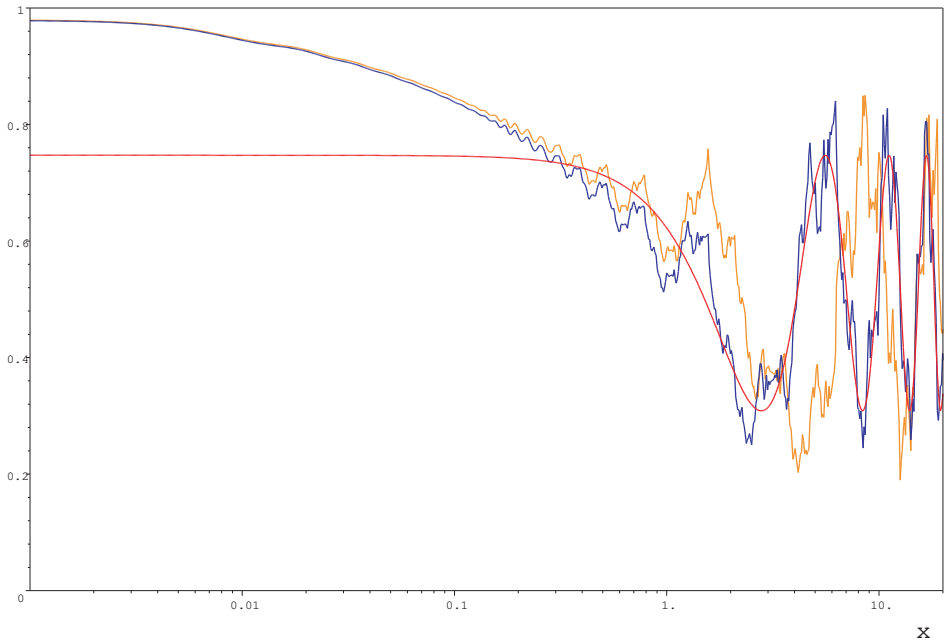}}
\caption{Comparison of the survival probability computed using the exact 
eigenvalues or the approximation $\lambda _n =n$ (highest curve to the left)
 for $\xi =1/3$. 
For small $x$, the exact survival probability gives lower values, while for $x>1$, the $L/E$ 
dependence can be completely inverted. The result including only the first oscillatory term 
is also shown (lowest curve to the left).}
\label{fig:probapprox}
\end{figure}
\end{center}

Their squares $\lambda _n^2$ give the frequencies appearing in Eq.~(\ref{anu1nu1}). For $n \gg
\xi^2$, the first correction to $\lambda _n^2$ is $2 \xi^2$, which is an irrelevant global phase.
This suggests a strong harmonic behavior of the survival probability, which would justify the use
of the approximation $\lambda _n \approx n$. Figure~\ref{fig:probapprox} shows however that this 
approximation is very bad indeed when the oscillation regime is entered and 
that the exact values of $\lambda _n$ should be used.

The envelop of the modal amplitudes (\ref{uon}) is a Lorentzian centered at $\lambda=0$ and of
width $\xi \sqrt{1+\pi^2 \xi^2}$. For $\xi \ll 1/\pi$, the width is nearly $\xi$; for $\xi \gg 1/
\pi$, it  approaches $\pi \xi^2$. As the Lorentzian falls down rapidly, the width indicates the
number of relevant modes according to $\xi$. For numerical calculations, 
we have checked that it is safe to cut the infinite sum
in Eq.~(\ref{anu1nu1}) after a few widths. Actually at small $\xi$, this means one could even
retain  only the mode $n=1$.

From Eq.~(\ref{pnu1nu1}) it is straightforward that the mean value\footnote{The mean value is
understood  here as the average over a large interval in $x$.} of  the survival probability and the
amplitude of its fluctuations are given by 
\begin{eqnarray}
\label{probmoy}
\left< P_{\nu _1 \nu _1} \right> &=& \sum _{n=0}^{\infty} \left( U_{0n}\right) ^4 ,\\
\label{fluctu}
\sigma ^2 \left( P \right) &=& \left( \sum _{n=0}^{\infty} \left( U_{0n}\right) ^4 \right) ^2 
- \sum _{n=0}^{\infty} \left( U_{0n}\right) ^8.
\end{eqnarray}

\begin{center}
\begin{figure}[ht]
\centerline{\includegraphics{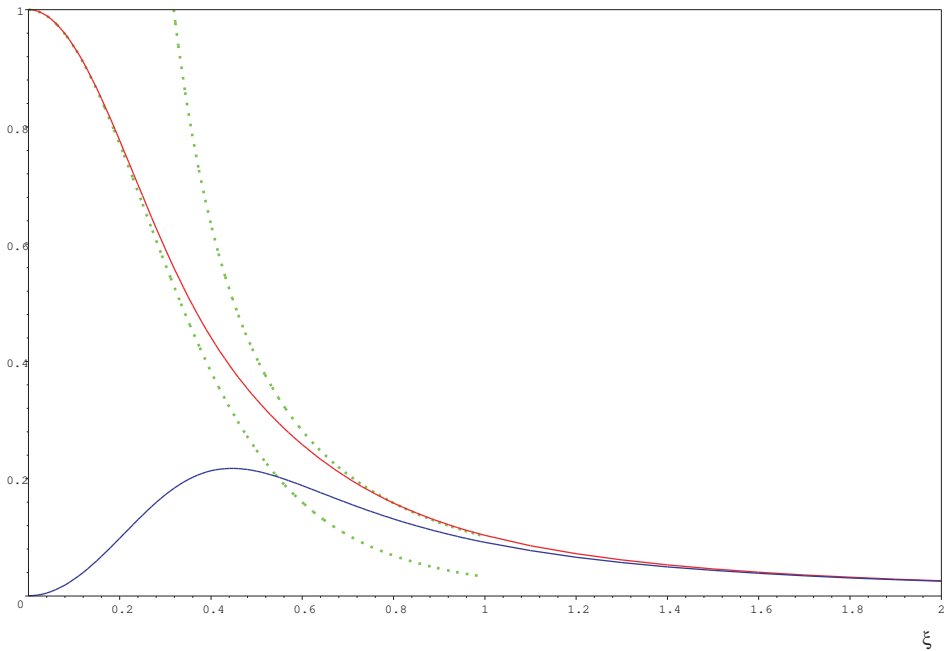}}
\caption{Mean survival probability $\left< P _{\nu _1 \nu _1} \right>$ and fluctuations $\sigma(P)$ 
as functions of $\xi$. 
The dashed lines give the small $\xi$ ($\left< P _{\nu _1 \nu _1} \right> \approx (U_{00})^4)$,
and the large $\xi$ ($\left< P _{\nu _1 \nu _1} \right> \approx \frac{1}{\pi^2 \xi^2}$) 
approximations.}
\label{fig:meanprob}
\end{figure}
\end{center}

These results are obtained without any approximation.
The mean value $\left< P_{\nu _1 \nu _1} \right>$ is dominated by the zero-mode contribution $\left(
U_{00}\right) ^4$ for $\xi \leq 1/3$, while the large $\xi$ regime, 
$\left< P_{\nu _1 \nu _1} \right> = \frac{1}{\pi^2 \xi^2}$
is entered from $\xi \approx 0.8$. 
At large $\xi$, the amplitude of the fluctuations $\sigma \left( P \right)$ 
tends asymptotically to $\left< P_{\nu _1 \nu _1} \right>$ (see Fig.~\ref{fig:meanprob}).

It is worth noting another interpretation of this result. Indeed, if we suppose instead that the
phases in (\ref{anu1nu1}) are uncorrelated, i.e., random frequencies, and perform an average 
on a set of phase values, we recover the expressions (\ref{probmoy},\ref{fluctu}). This means the
frequencies in Eq.~(\ref{anu1nu1}), although quasiharmonic, do not impose a periodic behavior to
the survival probability, especially at large $\xi$. The only harmonic leftover consists in very
narrow (and thus phenomenologically irrelevant) periodic peaks in $\left< P_{\nu _1 \nu _1}
\right>$.

Briefly, at large $x$, the survival probability $\left< P_{\nu _1 \nu _1} \right>$ has two regimes.

For small $\xi$, the result is dominated by a constant $\left(U_{00}\right) ^4$, to
which an oscillatory term of frequency $\left( \lambda _1^2 - \lambda _0^2 \right) \approx 1$ is
added. 

At large $\xi$,
the survival probability drops quickly to reach its large $x$ regime: a high frequency
fluctuation around its mean value $\frac{1}{\pi^2 \xi^2}$. From a physical point of view, 
in both cases it is safe to average the survival probability to its mean
value if the size of the detector or the source is large (typically $\Delta x \gg 1$).
The shape of the survival probability for typical values of $\xi$ is shown on Fig.~\ref{fig:prob}. 

\subsection{Experimental constraints}

We try here to summarize the main experimental results for the survival
probabilities $P_{\nu _e \nu _e}$ and $P_{\nu _\mu \nu _\mu}$. For $\nu _e$
we discuss the CHOOZ experiment, the solar, and the atmospheric neutrinos; for
$\nu _\mu$, the KARMEN and LSND experiments, the K2K experiment, and the
atmospheric neutrinos.

We first discuss the constraints on $P_{\nu _e \nu _e}$.
The CHOOZ experiment\cite{chooz}
gives a very clear and strong constraint on $P_{\nu _e \nu _e}$. Clear, because
the $L/E$ ratio spans less than one order of magnitude. Strong, because the $L/E$ ratio
is large (actually larger than for the K2K experiment) and the experiment does not
observe any suppression with high precision: $P_{\nu _e \nu _e} > 0.95$, at 90\% C.L.

The solar neutrinos are observed by many experiments. 
We list the observed fluxesin Table~\ref{table}.

\begin{table}[htbp]
\label{table}
\caption{Observed fluxes of the solar neutrinos.}
\begin{center}
\begin{tabular}{cccc} 
Experiment & Observed to expected & Experimental & Theoretical\\
 & flux ratio & errors & uncertainty (within SSM) \\ \hline
SuperKamiokande \cite{SKsolar} & 0.47 & $\pm$ 0.02 & +0.09 / -0.07 \\ 
Gallex/GNO/Sage\cite{GAsolar} & 0.60 & $\pm$ 0.06 & +0.04 / -0.03 \\ 
Homestake \cite{CLsolar} & 0.33 & $\pm$ 0.03 & +0.05 / -0.04 \\
\end{tabular}
\end{center}
\end{table}

The expected fluxes and theoretical uncertainties are taken from the 1998
Bahcall-Basu-Pinsonneault (BBP98) standard solar model \cite{BP98} (SSM). The highest experimental
precision is reached by SuperKamiokande\cite{SKsolar}. We however stress that a sizable uncertainty
still lies in the nuclear cross section\cite{Bcross} of the reaction $^7{\rm Be} +p \rightarrow \;
^8 \! {\rm B} + \gamma$, resulting in a 20\% uncertainty in SuperKamiokande. Therefore, the gallium
experiments and SuperKamiokande together could be accounted for by a global suppression (neutrino
disappearance of 40 to 60\%, within the SSM uncertainty). Homestake has a lower value, but is in
addition sensitive to the Be lines, for which direct information will only be provided by the
upcoming SNO and Borexino experiments.

SuperKamiokande searched for but did not observe any other effect at its current sensitivity:
no spectral distortion\footnote{Except at the largest energies, possibly explained by the
so-called $^3{\rm He} +p$ (hep) neutrinos\cite{hep}.}, no seasonal effect, no day-night effect. 
A global, energy and distance independent suppression is thus not an unreasonable 
explanation of the observed solar neutrino deficits.

The atmospheric $\nu _e$ flux, observed by
SuperKamiokande\cite{SKatmos}, does not show any distance or energy dependent suppression. The
observed angular and energy dependence agrees with the expected one. The main uncertainty here lies
in the absolute flux expectation. It turns out that at most 20 to 30\% average suppression, without angular or energy dependence,
is allowed by the combination of data and Monte Carlo calculations \cite{atmosfluxes}. 

More robust is the value of the double ratio 
\begin{eqnarray*}
R= \frac{\left( \nu _\mu / \nu _e \right)_{\rm observed}}
{\left( \nu _\mu / \nu _e \right)_{\rm expected}} \approx \frac 23.
\end{eqnarray*}

We now turn to the discussion of the constraints on $P_{\nu _\mu \nu _\mu}$.
The K2K experiment will soon give a constraint on $P_{\nu_\mu \nu_\mu}$ almost as clean and strong
as what CHOOZ obtained for the $\nu _e$ survival probability. 
The $L/E$ ratio also spans less than one order of magnitude. The last
available information \cite{K2K} quotes the following result: 27 observed events, 40$\pm$5 expected.

The atmospheric $\nu _\mu$ flux, observed in detail by SuperKamiokande, shows a strong
angular and energy dependent suppression.
The observed flux, plotted against $L/E$, shows a decrease from $P=1$ to $P=\sim 0.6$, 
where it levels off. These values are tainted by large errors\footnote{The
recoil spectrum washes out the sensitivity to the survival probability.} and are not corrected to
take into account initial flux uncertainties (which could be judged from the $\nu_e$'s), and thus an
extra overall suppression of 20 to 30\% is not excluded. Note that the $L/E$ ratio, similar for
the $\nu _e$'s and $\nu _\mu$'s, spans many orders of magnitude, and  therefore provides a strong
constraint, as we will see in the next subsection.

The KARMEN experiment \cite{karmen} quotes negative results, i.e., it does not see any
$\bar{\nu}_\mu$ transition to $\bar{\nu}_e$ at small $L/E$ ratio. An upper limit of $6.5 \times
10^{-4}$ is put on the transition probability. On the contrary, the LSND experiment \cite{lsnd}
claims a 0.3\% transition probability $P_{\nu _e \nu _\mu}$.

\subsection{Comparison with experimental data}

How can these data be accounted for by the simplest model (one $\nu _L$, one KK tower)?

For $\nu _e$, we are looking here for possible across the board suppressions at large $x$ 
by a factor of 40 to 60\%, such a solution is the simplest interpretation of the absence 
of $L/E$ dependence in the SuperKamiokande solar neutrinos data. 
To prevent any $L/E$ dependence, we should also avoid MSW effects in the Sun or Earth.

As the Sun-Earth system is a very long-baseline system and as the solar core is large 
(typically, $x \sim 10^5$ and $\Delta x \sim 10^2 \gg 1$ for solar neutrinos 
with $E \sim 1$ MeV and $R \sim 1$ mm),
the only observable effect will be an average suppression. If we fix the suppression range from 40
to 60\%, i.e., the mean value $\left< P_{\nu_e \nu_e} \right>$ at large $x$, we have to take
$0.29< \xi < 0.42$ approximately. At fixed $\xi$, we then extract the largest allowed $x$
position for CHOOZ still fitting the data. 
As $L/E$ is known for CHOOZ, this is nothing but an upper limit for $1/R^2$.

\begin{center}
\begin{figure}[ht!]
\centerline{\includegraphics{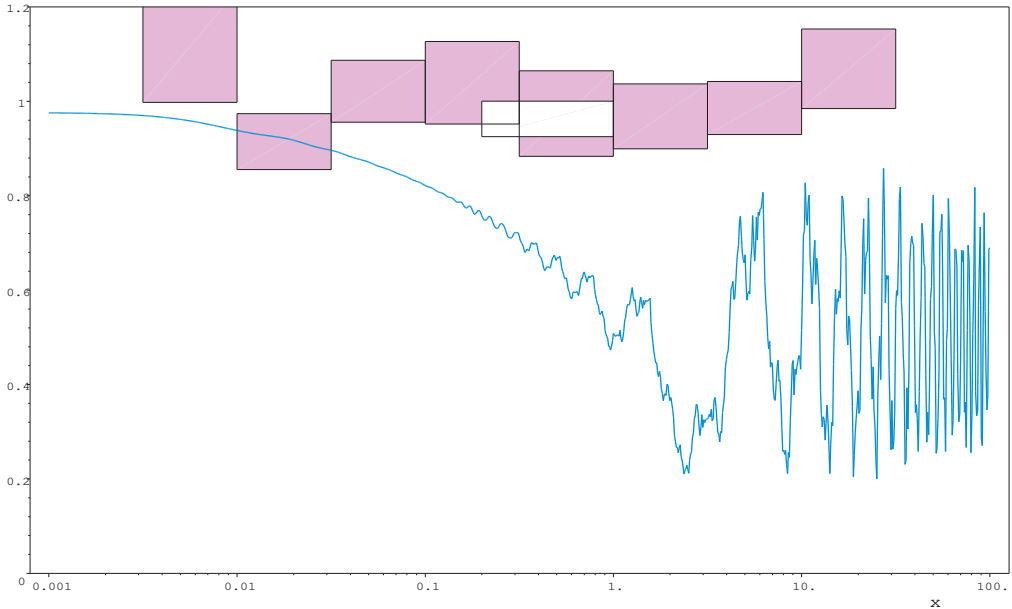}}
\caption{Comparison of the toy model with the $\nu _e$ constraints. The mean survival probability
at large $x$ is fixed at 50\% to fit the solar neutrinos. We have plotted the latest available
experimental data we could find. The series of filled boxes show the $L/E$ dependence for the
atmospheric $\nu _e$, as observed by SuperKamiokande (1$\sigma$). The data has been normalized by
an overall 0.95 factor with respect to the raw data (instead of the usual $\sim$ 0.9 used by
SuperKamiokande). The open box corresponds to the CHOOZ constraint. The error bar at $2\sigma$
level combines quadratically the statistical and systematic errors (see~[10]). The boxes
cannot slide to the left without getting into the range of the MSW effect, showing thus a complete
disagreement with the toy model.}
\label{fig:nue11}
\end{figure}
\end{center}

As $1/R^2$ never exceeds $10^{-5}{\rm eV}^2$, we always lie in the MSW mass range\footnote{Even if
we do not exclude {\it a priori} the MSW effect, due to the ''large'' value of $\xi$, many states of
the tower participate, contrary to a few ones in \cite{dvali}. A rough evaluation, using the same
method as in \cite{dvali}, shows no nonadiabatic rise at the largest energies.}. 
The absence of MSW free solution is clearly depicted in Fig.~\ref{fig:nue11}.
A fit to the solar neutrino problem including the MSW effect has first been proposed
in~\cite{dvali}. It needs a very small coupling constant $\xi \sim 10^{-3}$ and gives 
an energy dependent suppression which can fit the SuperKamiokande results only due to 
the limited energy resolution of the detector.

We look however for a more drastic, energy independent suppression, which thus needs 
to avoid MSW effects. This could be reached in two ways. 
The first consists in increasing the mass of the neutrinos by a constant number,
so as to exceed the MSW threshold. 
Such an approach could be investigated along the lines of~\cite{ramond} but will 
not be pursued here. Another possibility consists in extending our model to
two active neutrinos coupled to one KK tower.

\begin{center}
\begin{figure}[ht!]
\centerline{\includegraphics{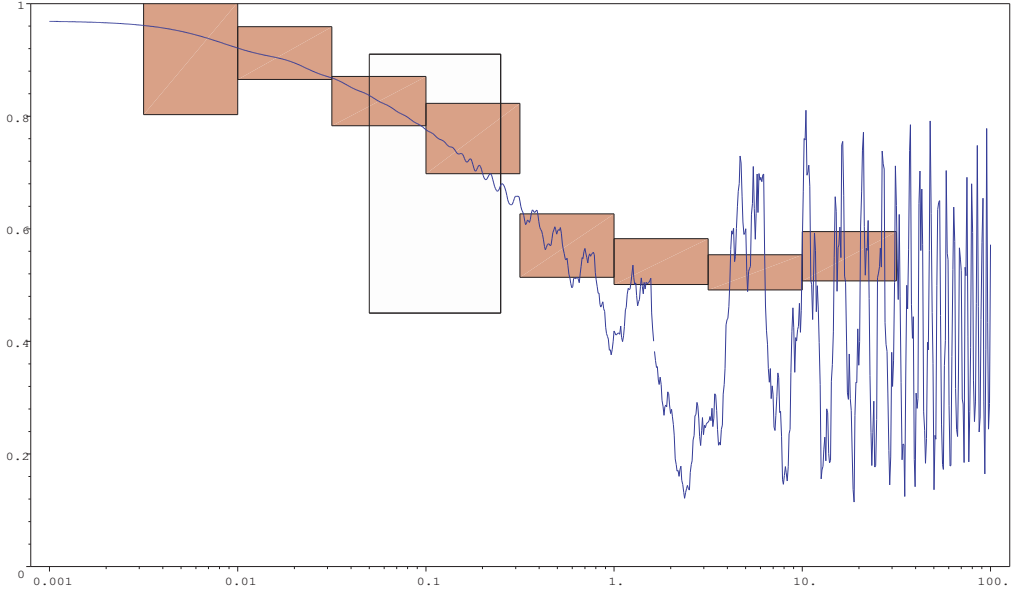}}
\caption{Comparison of the toy model with the $\nu _\mu$ constraints ($\xi=4/10$).
The series of filled boxes show the $L/E$ dependence for the atmospheric $\nu _\mu$, 
as observed by SuperKamiokande (1$\sigma$; the same 0.95 factor has been used). 
The open box corresponds to the K2K constraint. The error bar at $1\sigma$ level
combines quadratically the statistical and systematic errors. 
The survival probability is in good agreement with
the constraints, once an average on each energy "bin" is performed.}
\label{fig:numu11}
\end{figure}
\end{center}

One could also ask whether it would be possible to fit the $\nu _{\mu }$ 
constraints in the toy model (that is $\nu _1=\nu _\mu$). 
Figure~\ref{fig:numu11} shows that a good agreement with the experimental data can be obtained 
for $\xi=0.4$.

Thus far, we have dealt only for atmospheric neutrinos with the vacuum propagation,
neglecting the Earth density effect. This is correct for low energy neutrinos (say,
less than 5 GeV for a $\delta m^2$ of $10^{-3} {\rm eV}^2$), but
must be revised for high energy particles where the resulting effective potential
becomes determinant. In fact, this is the main reason why sterile neutrinos
are now disfavored by the SuperKamiokande experiment in the usual (purely
four-dimensional) case \cite{sk_sterile}. Namely, while a small $\delta m^2$ guarantees
the desired oscillations and large mixing angle in vacuum, the large effective mass difference
between $\nu_s$ and $\nu_e$ or $\nu_\mu$ due to the Earth's presence lifts the near degeneracy
and in practice blocks the oscillations when the density is sufficient.
In the four-dimensional case, this occurs for both $\nu$'s and $\bar \nu$'s, and results in
a contradiction with high energy data (upward through-going muons and partially contained events).
Another but rather weaker constraint comes from the multiring, NC enriched sample.

What is the situation in the extra-dimensional case?

We differ from the standard case by the presence of a tower of states. This
means that, whatever the density, high energy $\bar \nu$'s will always find a state
to oscillate to.
The effect will thus be at most 1/2 of that expected in the standard situation,
with little suppression in the $\nu$'s and large suppression in the $\bar \nu$'s.
We are thus confident that no serious contradiction occurs.

A detailed simulation of the Earth and angular acceptances is beyond the scope of
this paper and will be attempted when we have explored further the full parameter
space (which may include additional mass terms for the bulk and/or brane fermions~\cite{ramond2}).

\section{Two neutrinos coupled to one bulk fermion: the 2-1 model}
\label{s:sec3}

\subsection{Formalism}

To solve the solar neutrino puzzle, we now extend the toy model
to include a second neutrino in four-dimensional space. 
We will call $\nu _e$ and $\nu _f$ the two neutrinos 
states living in 4 dimensions (we will later discuss the
possibility $\nu _{f }=\nu _{\mu }$). The coupling of the flavor neutrinos to the
bulk neutrino is chosen in a particularly economical way, the Lagrangian being 
\begin{equation}
{\cal L} =\lambda _{e}\overline{\nu }_{e}\Psi (x^{\mu },y=0)H(x^{\mu
})+\lambda _{f }\overline{\nu }_{f }\Psi (x^{\mu },y=0)H(x^{\mu }) .
\end{equation}

Therefore, from the four-dimensional point of view, only one linear combination
of the two flavor neutrinos, which we call as previously $\nu _{1}$, will be coupled to the
Kaluza-Klein tower associated with the bulk fermion $\Psi $. Namely, 
\begin{eqnarray}
\nu _{1} &=&\cos \theta \;\nu _{e}+\sin \theta \;\nu _{f } ,\cr
\nu _{2} &=&-\sin \theta \;\nu _{e}+\cos \theta \;\nu _{f },
\end{eqnarray}
with $\cos \theta =\frac{m_{e}}{m}$ , $\sin \theta =\frac{m_{f }}{m}$, $m=%
\sqrt{m_{e}^{2}+m_{f }^{2}}$ and $m_{e,f }=\frac{\lambda _{e,f }v}{%
\sqrt{2\pi R}}$ \footnote{Here, $m_{e}$ and $m_{f}$ are simply mass parameters
without any link to the charged fermion masses.}. 
The orthogonal combination $\nu _{2}$ remains
massless.While the mixing with bulk states proceeds exactly as before,
 the physical consequences can be
quite different. Indeed, a new phenomenological parameter, 
the mixing angle $\theta $ now plays a crucial role
in the survival probabilities of the flavor neutrinos,

\begin{eqnarray}
P_{\nu _{e}\nu _{e}} &=&\cos ^{4}\theta \;P_{\nu _{1}\nu _{1}}+\sin
^{4}\theta +2\sin ^{2}\theta \;\cos ^{2}\theta \;\textrm{Re}\left( A_{\nu
_{1}\nu _{1}}\right) ,\cr
P_{\nu _{f }\nu _{f }} &=&\sin ^{4}\theta \;P_{\nu _{1}\nu _{1}}+\cos
^{4}\theta +2\sin ^{2}\theta \;\cos ^{2}\theta \;\textrm{Re}\left( A_{\nu
_{1}\nu _{1}}\right).
\end{eqnarray}
The flavor transition probability is given by 
\begin{equation}
P_{\nu _{e}\nu _{f }}=P_{\nu _{f }\nu _{e}}=\sin ^{2}\theta \ \cos
^{2}\theta \ \left[ P_{\nu _{1}\nu _{1}}-2\textrm{Re}\left( A_{\nu _{1}\nu
_{1}}\right) +1 \right] .
\end{equation}

This model (hereafter called the 2-1 model) has 3 degrees of freedom, 
$(\xi ,\theta ,R)$, to fit the
experimental data which also include the different bounds to the $\nu _{\mu
}(\overline{\nu }_{\mu })$ fluxes, in case $\nu _{f}=\nu _{\mu}$ is chosen. 
We should also stress that $\theta$ is not the $\nu _e - \nu _f$ mixing angle. Such mixing 
arises, but only as a result of the independent coupling of both states to the bulk neutrino.

\subsection{Comparison with experimental data}

In the first part of this analysis, we will concentrate on the electronic
neutrino data. We will show that, taking advantage of the second active neutrino, 
we can now fit all the experimental constraints on the electronic neutrino 
without getting into the MSW region.

\begin{center}
\begin{figure}[ht!]
\centerline{\includegraphics{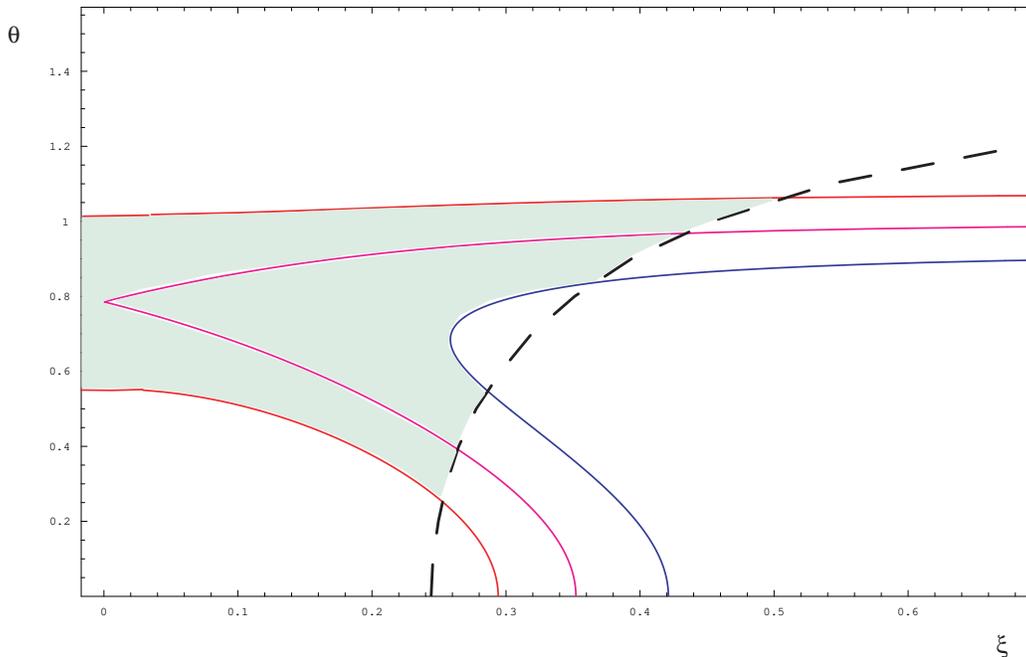}}
\caption{Region of values for $\xi$ and $\theta$ for which the solar 
and CHOOZ constraints for $\nu _e$ can be accommodated without MSW effect.
The solid lines correspond (from top to bottom) to a solar neutrino mean survival 
probability of 60, 50, and 40\%. The dashed line 
indicates the frontier of the MSW region in view of the CHOOZ constraint. 
The filled region gives the allowed $(\xi,\theta)$ values.}
\label{fig:theta}
\end{figure}
\end{center}

The solar neutrino deficit is accounted for if the $\nu _{e}$ mean survival 
probability at large $x$ ranges between 40\% to 60\% max. as a 
result of the experimental and SSM uncertainties. This constraint
defines a region in the plane $\xi -\theta $ as shown in Fig.~\ref{fig:theta}. 

For small values of $\theta $, $\nu _{e}$ is mainly $\nu
_{1}$ , so that the allowed values of $\xi $ range from 0.3 to 0.42. 
For higher values of $\xi $ indeed, the electronic neutrino survival
probability drops under 40\%, while for lower values, it does not decrease below 
60\%. As the mixing angle $\theta $ increases, the proportion of the
decaying component of $\nu _{1}$ diminishes, so that the allowed range for $%
\xi $ gets enlarged. Finally, if $\theta $ is too big ($1<\theta <\pi /2$
approximately), the decaying component of $\nu _{e}$ is insufficient to
explain the solar neutrino deficit.

The next constraint comes from the CHOOZ nuclear reactor
experiment. As seen before, the $\overline{\nu }_{e}$ produced in the reactor 
with a typical energy of 2 MeV show no disappearance at a distance $L=1~$ km. 

For given $\xi$ and $\theta$, a maximum admissible value of $x$,
 or equivalently, a minimum value 
of $R$ (the radius of the compactified extra dimension) results. 
The value of $x$ increases with $\theta $ at given $\xi $, but decreases with $\xi $
at fixed $\theta $ . Therefore, a small coupling constant $\xi$ and 
a large mixing angle $\theta$ are preferred in view of the CHOOZ
experiment.

On the other hand, $1/R$ controls the typical mass difference between two
consecutive Kaluza-Klein levels. Therefore, MSW resonant conversion will
take place if $1/R$ is of the same magnitude order as the MSW potential. 
As suggested by the latest SuperKamiokande data, we can put an upper bound on $R$
and avoid the MSW effect. 
Typically, $R_{\max }\simeq 10^{-2}$ mm. As a result, for some $\xi$ and $\theta$,
this bound can be incompatible with the CHOOZ constraint. This is shown in Fig.~\ref{fig:theta}. 
The constraint arising from the unsuppressed flux of the atmospheric $\nu _e$ 
will be discussed later, as it depends on the flavor of the second active neutrino. 
We can already announce that it favors the case $\nu _{f}=\nu _{\mu}$, yet the case
$\nu _{f}=\nu _{\tau}$ remains also possible.
Therefore, all constraints on $P_{\nu _e \nu _e}$ can be satisfied in the 2-1 model. 
It is also clear that a nonzero mixing angle $\theta>0$ is needed.

\begin{center}
\begin{figure}[ht!]
\centerline{\includegraphics{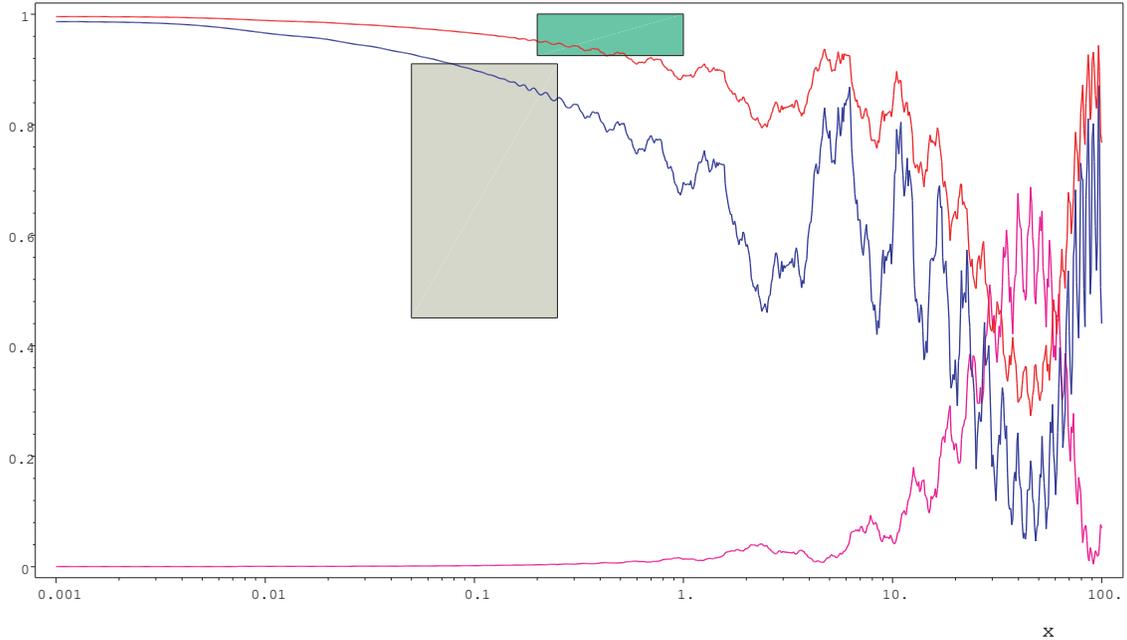}}
\caption{Comparison of the 2-1 model with the CHOOZ and K2K constraints.
(highest curve is for $\nu_e$)
$\xi=0.3$ and $\theta=1.05$, so that $\left<P_{\nu _e \nu _e} \right> \simeq 60\%$. 
The transition probability $P_{\nu _e \nu _\mu}$ is also depicted (lowest curve).}
\label{fig:nu21}
\end{figure}
\end{center}

We now study in detail the possibility $\nu _{f}=\nu _{\mu}$.
We first discuss the $\overline{\nu }_{\mu }$ disappearance experiment K2K, 
which reveals some 30\% deficit for 2 GeV neutrinos at a distance $L\simeq 250$ km. 
We have $x_{\rm K2K}\simeq 1/4 \cdot x_{\rm CHOOZ}$, so that muonic neutrinos are expected to
disappear more than electronic neutrinos. This requires $\theta >\pi /4$,
and higher values of $\xi $ are favored, as $P_{\nu _{e}\nu _{e}}-P_{\nu
_{\mu }\nu _{\mu }}\propto (1-P_{\nu _{1}\nu _{1}})$. However, even for the
maximal allowed $\xi $, the preliminary result of K2K can only be accommodated by taking
the large statistical error into account. Figure~\ref{fig:nu21} shows a possible fit for $\nu _e$ 
and $\nu _\mu$, which solves the solar neutrino problem, and simultaneously
satisfies the CHOOZ and K2K constraints. It will be shown hereafter that this fit also
satisfies the constraints for the atmospheric neutrinos. 

\begin{center}
\begin{figure}[ht!]
\centerline{\includegraphics{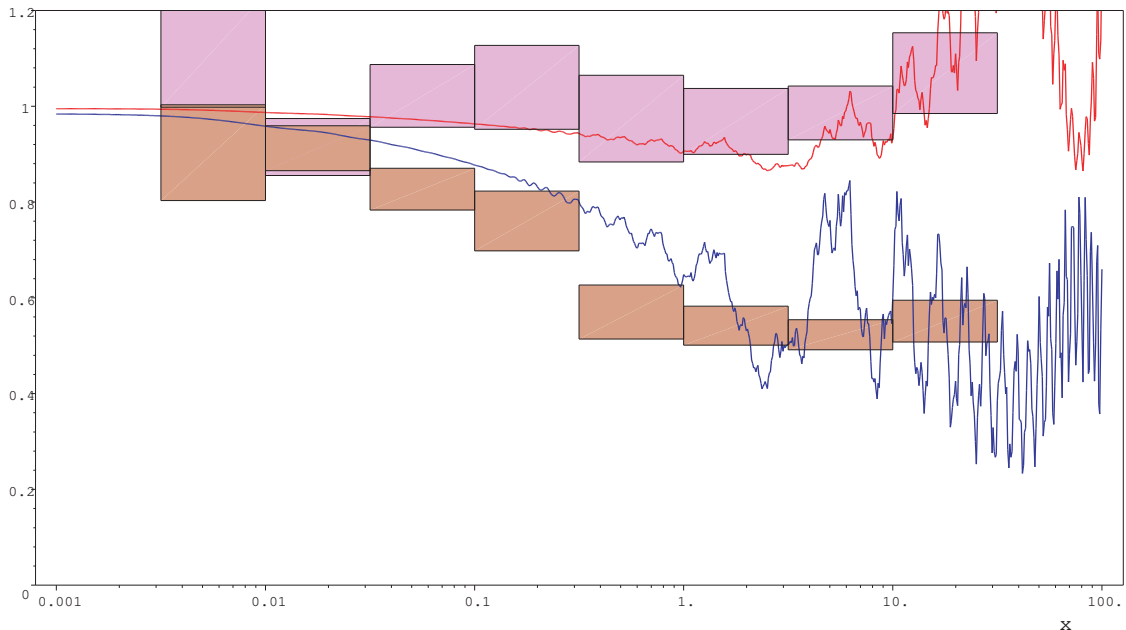}}
\caption{Expected atmospheric neutrino fluxes in the 2-1 model with $\nu _f=\nu _\mu$ 
  and the SuperKamiokande data. $\xi=0.3$ and $\theta=1.05$ as in
  Fig.~\ref{fig:nu21}. The initial flux $2 \stackrel {( - )}{\nu _\mu }+
  \stackrel {( - )}{\nu _e }$ is normalized to 1 at $x=0$. The observed
  SuperKamiokande data has been normalized as in Fig.~\ref{fig:nue11}. The
  agreement with experimental data is quite remarkable.}
\label{fig:nuSK21}
\end{figure}
\end{center}

To discuss the constraints coming from the atmospheric neutrinos, we recall that in the 
2-1 model, a transition $\nu _e \rightarrow \nu _\mu$ or $\nu _\mu \rightarrow \nu _e$ 
becomes possible. The transition probability, as shown in Fig.~\ref{fig:nu21}, is non-negligible
in the range of the atmospheric neutrinos. As the atmospheric neutrinos originate from the 
decay of the charged pions and kaons into muons and the subsequent decay of muons into  
electrons, the ratio of the neutrino initial fluxes 
$\frac{\phi ^{(i)}_{\nu _\mu}}{\phi ^{(i)}_{\nu _e}}$ 
is expected to be very close to 2, especially at low energy\footnote{At higher
energy, the produced muon can go through the atmosphere without decaying, so that the
ratio $\frac{\phi ^{(i)}_{\nu _\mu}}{\phi ^{(i)}_{\nu _e}}$ increases with energy.}. 
Therefore, the expected neutrino flux in the 
2-1 model with $\nu _f = \nu _\mu$ is given by (we do not distinguish between $\nu$ and
$\overline{\nu}$) \begin{eqnarray}
\frac{\phi _{\nu _e}}{\phi ^{(i)}_{\nu _e}}=P_{\nu _e \nu _e}+ 2 \, P_{\nu _\mu \nu _e} \\
\frac{\phi _{\nu _\mu}}{\phi ^{(i)}_{\nu _\mu}}=P_{\nu _\mu \nu _\mu}
+ 1/2 \, P_{\nu _e \nu _\mu},
\end{eqnarray}
As a result, the observed $\nu _e$ flux can be enhanced compared to the initial production 
flux in our model. In Fig.~\ref{fig:nuSK21}, we see that this picture is in very good agreement with 
the SuperKamiokande results. 

We are left with the constraints of KARMEN and LSND. 
The negative result of the KARMEN experiment can easily be accommodated as 
$x_{\rm KARMEN}\simeq 3\times 10^{-4}\ x_{\rm CHOOZ}$. 
On the contrary, as $x_{\rm LSND}\simeq 6\times 10^{-4}\ x_{\rm CHOOZ}$, our model can never 
comply with the LSND results, for any allowed values of $\theta $ and $\xi $.

We have thus shown that all experimental data (with the exception of LSND) 
can be accommodated in the simple 2-1 model with $\nu _f = \nu _\mu$,that 
is with $\nu _e$ and $\nu _\mu$ coupled to the same Kaluza-Klein tower. 
However the fit could be invalidated in the near future should the LSND 
signal be confirmed by an independent experiment. A critical test will also 
be provided by the improving accuracy of the K2K experiment. We also point out 
that the astrophysical bound could be evaded in this model, since the disappearance 
of $\nu _e$ or $\nu _\mu$ in the extra dimensions is never complete (see~\cite{dienes}).

Can we improve the fit to $\nu _e$ data by allowing for $\nu _f \neq \nu _\mu$ ? 
For instance, we could have $\nu _f = \nu _\tau$ or $\nu _f=\nu _s$, 
with $\nu _s$ an extra four-dimensional sterile neutrino. Of course, the model 
is no longer affected by $\nu _\mu$ observations, but strong constraints like 
CHOOZ remain, and must be reconciled with solar data, given the allowed 
region of Fig.~6. The real difficulty could, however, come from the 
atmospheric $\nu _e$. In the ($\nu _e - \nu _\mu - KK$) model indeed, 
the atmospheric  $\nu _e$ flux at long distance was boosted by a 
$\nu _\mu \rightarrow \nu _e$ conversion. Here, with the $\nu _e$ 
and $\nu _\mu$ sectors separated, boosting disappears, and the $\nu _e$ 
flux falls more rapidly. Still, given the uncertainty on the absolute 
normalization of the atmospheric $\nu _e$, no contradiction exists for the 
moment. Actually, the constraint provided by the CHOOZ experiment appears 
to be more severe, so that the allowed region of parameters that give 
a good fit to the $\nu _e$ data is still given by Fig.~6. 

We have also checked that a ($\nu _e - \nu _\mu - \nu _\tau - KK$) 
approach along the present lines does not improve the quality of the fit. 

\section{Conclusions}
\label{s:concl}

We have shown that a simple model with 2 massless neutrinos coupled to 
one Kaluza-Klein tower meets most experimental constraints (except for 
LSND), and differs from the oscillation image by the energy dependence of 
the neutrino disappearance. This model can be developed by adding extra 
parameters in the form of bare masses for the neutrinos, while simply 
increasing the number of neutrinos coupled to the Kaluza-Klein states 
brings little gain.

\section*{Acknowledgments}

This research was initiated through early discussions with Dr. Patricia Ball, who we 
want to thank here.

This work was partially supported by the I.~I.~S.~N. (Belgium), and by the 
Communaut\'e Fran\c caise de Belgique - Direction de la Recherche Scientifique 
programme ARC.

Y.~Gouverneur benefits from an A.~R.~C. grant.
F.-S.~Ling benefits from a F.~N.~R.~S. grant.
D.~Monderen and V.~Van~Elewyck benefit from a F.~R.~I.~A. grant.

\end{document}